\documentclass[twocolumn,pra,aps,superscriptaddress,english,floatfix]{revtex4-2}
\usepackage{amsmath}
\usepackage{amssymb}
\usepackage{amsfonts}
\usepackage[pdftex]{graphicx}
\usepackage{subfigure} 
\usepackage[english]{babel}
\usepackage{braket}
\usepackage{color}
\usepackage{mathtools}
\usepackage[super]{nth}
\usepackage{threeparttable}
\usepackage[colorlinks,linkcolor=blue,anchorcolor=blue,citecolor=blue,urlcolor=blue]{hyperref}
\usepackage{babel}
\makeatother

\begin{document}

\preprint{APS/123-QED}

\title{Wavelength-scale Optical Parametric Oscillators}

\author{Saman Jahani}%
\author{Arkadev Roy}%
\author{Alireza Marandi}%
\email{marandi@caltech.edu}
\affiliation{%
Department of Electrical Engineering, California Institute of Technology, Pasadena, CA 91125, USA.
}%

\date{\today}

\begin{abstract}
Despite recent progress in nonlinear optics in wavelength-scale resonators, there are still open questions on the possibility of parametric oscillation in such resonators. We present a general approach to predict the behavior and estimate the oscillation threshold of multi-mode subwavelength and wavelength-scale optical parametric oscillators (OPOs). As an example, we propose an OPO based on Mie-type multipolar resonances, and we demonstrate that due to the low-Q nature of multipolar modes in wavelength-scale resonators, there is a nonlinear interaction between these modes. As a result, the OPO threshold, compared to the single-mode case, can be reduced by a factor which is significantly larger than the number of interacting modes. 
The multi-mode interaction can also lead to a phase transition manifested through a sudden change in the parametric gain as well as the oscillation threshold which can be utilized for enhanced sensing.
We establish an explicit connection between the second-harmonic generation efficiency and the OPO threshold. This allows us to estimate the OPO threshold based on measured or simulated second-harmonic generation in different class of resonators, such as bound-state in the continuum and inversely designed resonators. Our approach for analyzing and modeling miniaturized OPOs can open unprecedented opportunities for classical and quantum nonlinear photonics.

\end{abstract}

\maketitle

\section{Introduction}

Optical parametric oscillators (OPOs) have been widely used for may applications ranging from metrology and spectroscopy to quantum information science \cite{eckardt1991optical, schliesser2012mid, breunig2016three, marandi2012coherence, kippenberg2004kerr, okawachi2015dual, inagaki2016large,muraviev2018massively,roslund2014wavelength}. OPOs consist of a medium with quadratic or Kerr nonlinearity within a resonator, which is typically much larger than the operation wavelength, converting pump photons to signal and idler photons \cite{schliesser2012mid, marandi2012coherence, kippenberg2004kerr, okawachi2015dual, inagaki2016large, breunig2016three}. 
At degeneracy, the indistinguishable signal and idler of an OPO can form a squeezed vacuum state below the oscillation threshold \cite{milburn1981production, wu1986generation} that have been used for several applications in quantum information processing \cite{roslund2014wavelength, chen2014experimental, morin2014remote, nehra2019state}. Above threshold, the conversion efficiency boosts rapidly and the output signal illustrates a binary phase state which can be utilized as a spin in an  artificial Ising network \cite{marandi2014network, mcmahon2016fully}. Above-threshold degenerate OPOs have also been effectively used for generation of mid-IR frequency combs \cite{marandi2012coherence, muraviev2018massively, okawachi2020demonstration}.

Recent progress in nanoscale light confinement as well as precise nanofabrication of challenging nonlinear materials \cite{wang2018ultrahigh,lukin20204h} have inspired the idea of rethinking the possibilities of miniaturization of nonlinear systems to their extreme.  Miniaturized OPOs have recently been demonstrated in on-chip OPOs based on Kerr \cite{okawachi2015dual, kippenberg2004kerr, conti2004optical} and quadratic \cite{bruch2019chip} nonlinearities as well as whispering-gallery resonators \cite{werner2012blue}. The size of these resonators are still orders of magnitude larger than their operating wavelengths. Strong field confinement inside nanostructures has shed light on the possibility of nonlinear optics at nano-scale \cite{smirnova2016multipolar, jahani2014transparent, monticone2014embedded, nielsen2017giant, reshef2019nonlinear,yang2019high}. 
However, the main focus so far has been devoted to up-conversion in nanostructures, while optical parametric oscillation in wavelength-scale structures is still unexplored. The conventional theories which have mostly been developed for travelling wave nonlinear optical systems \cite{hamerly2016reduced} or high-Q resonators \cite{herr2014temporal,de2006multimode} cannot be directly applied to accurately model OPOs in nano-structures. The reason is that the spatial variation of the field happens in subwavelength regime where slowly-varying envelope approximation (SVEA) is not valid anymore \cite{hamerly2016reduced}. 
Moreover, unlike the conventional large-scale OPOs, in nano-structured resonators, the input pump can excite several modes of the cavity at the pump wavelength, and due the low-Q nature of modes, the pump can also directly interact with several modes at the signal wavelength.
Few theoretical models have been proposed recently to explain the spontaneous down-conversion in Mie resonators \cite{poddubny2018nonlinear, nikolaeva2020directional} and the threshold in 2D materials-based OPOs \cite{ciattoni2018phase}. However, these theories are either limited to specific structures or cannot explain the behavior of the system above the threshold.

\begin{figure}
\centering
\begin{tabular}{cc}

\includegraphics[width=8.5cm]{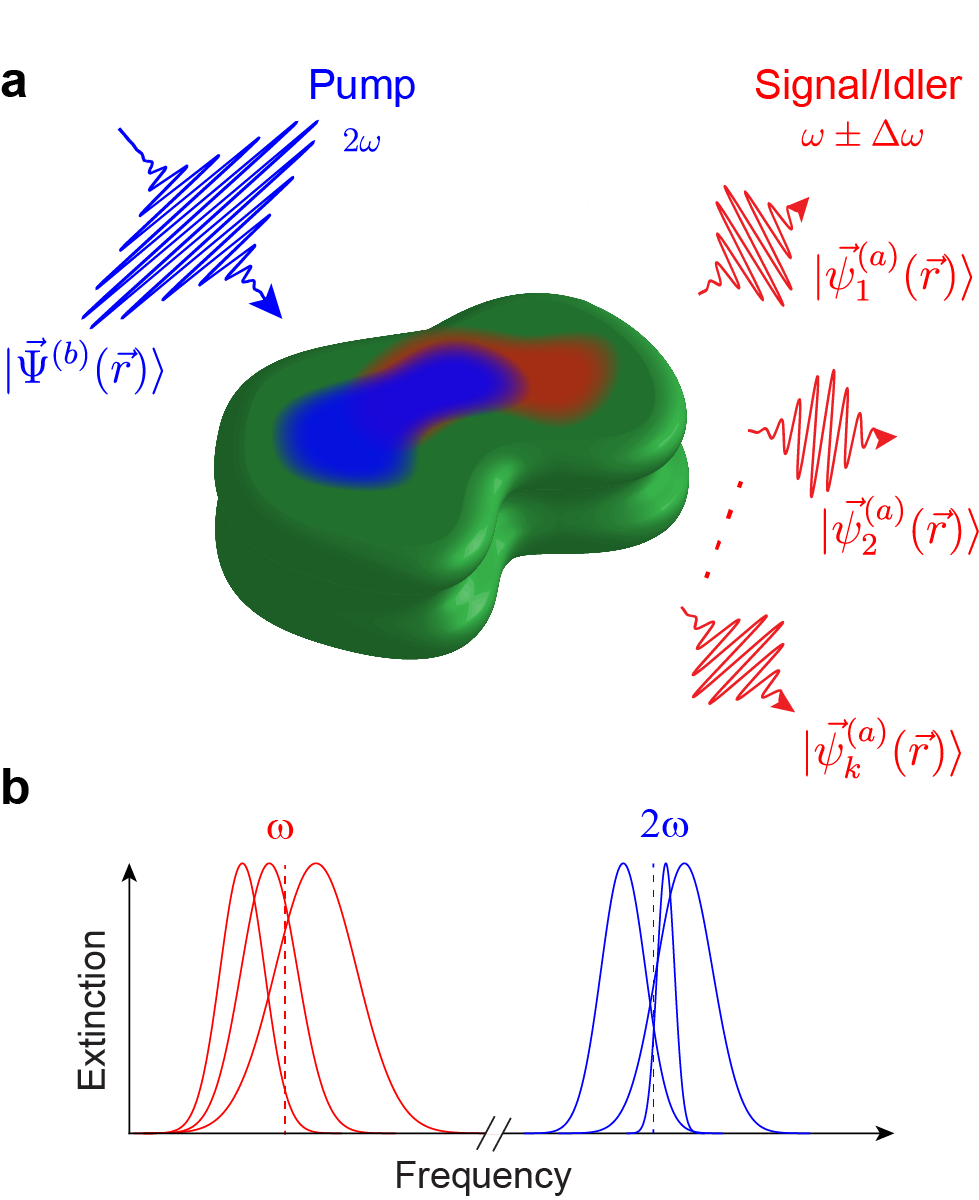}

\end{tabular}
\caption{{\bf Wavelength-scale optical parametric oscillators (OPOs)}. {\bf a,} An OPO with arbitrary geometry which resonates around the pump frequency ($2\omega$) and the half-harmonic ($\omega \pm \Delta\omega$). {\bf b,} The nonlinear behavior of the OPO can be determined by knowing the spatial overlap between the pump excitation at $2\omega$ and eigenmodes of the cavity around $\omega$ as well as the linear properties of the cavity around the pump and signal frequencies.} 
\label{fig:NanoOPO_Schematic}
\end{figure}

Here, we derive general conditions of parametric oscillation in subwavelength and wavelength-scale resonators. In the low-Q regime of these resonators, multiple modes around the signal wavelength can spectrally and spatially overlap (Fig.~\ref{fig:NanoOPO_Schematic}). This allows them to nonlinearly interact with each other through the pump. As an example, we estimate the OPO threshold in an AlGaAs nanoparticle which supports Mie-type multipolar resonances. We show that the multi-mode interaction at the signal wavelength can lead to a significant reduction in the threshold by a factor which is remarkably higher than the number of modes. The multi-mode interactions also result in a phase transition from degenerate to non-degenerate in these resonators with an abrupt change in the parametric gain and/or oscillation threshold which can be utilized for ultra-sensitive measurements. Moreover, we establish a connection between up-conversion processes in nanostructures and parametric down-conversion. This allows us to explore the possibility of OPO in the existing structures which have been offered for sum-frequency/second-harmonic generation. Our approach is general and can predict optical parametric oscillation in a wide range of nanostructured resonators, such as bound state in continuum, photonic crystal, and inversely designed cavities. 

\section{Theory}

To estimate the OPO threshold in multi-mode wavelength-scale resonators, we expand the field inside the cavity in terms of orthogonal eigenmodes (Fig.~\ref{fig:NanoOPO_Schematic}a), and we approximate the nonlinear dynamics of the electric field with a slowly varying envelope evolving in time-domain (see the Supplementary Material for more details). The electric field for the signal, idler, and pump can be expanded as the superposition of the quasi-normal modes as $\vec{E}(\vec{r},t) = \mathcal{E}_a\sum_k a_k(t)e^{-i(\omega-i\frac{\alpha_k}{2})t}|\vec{\psi}_k(\vec{r})\rangle+c.c.$, where $a_k$ is the slowly varying envelope \cite{haus1984waves, fabre1990squeezing, rodriguez2007chi}, $\mathcal{E}_a$ is the normalization constant such that $|a_k|^2$ is the energy stored in the k\textsuperscript{th} mode of the cavity, and for a homogeneous resonator, it is $\mathcal{E}_a=\sqrt{2/\varepsilon_0 n(\omega)^2}$, $|\vec{\psi}_k(\vec{r})\rangle$ is the cavity quasi-normal modes normalized such that $\langle{\vec{\psi}_m(\vec{r})\vec{\psi}_k(\vec{r})}\rangle=\delta_{mk}$ ($\delta_{mk}$ is the Kronecker delta), $\omega$ is the angular frequency of the signal ($\omega_s$), idler ($\omega_i$) or pump ($\omega_p$), $\alpha_k=\omega_k/Q_k$ is the decay rate of the cavity mode, $\omega_k$ is the eigenfrequency of the k\textsuperscript{th} mode with a quality factor of $Q_k$, and $\delta\omega_k=\omega-\omega_k$ is the detuning of the center of resonance of k\textsuperscript{th} from the frequency of the electromagnetic field. 

\begin{figure}
\centering
\begin{tabular}{cc}

\includegraphics[width=8.5cm]{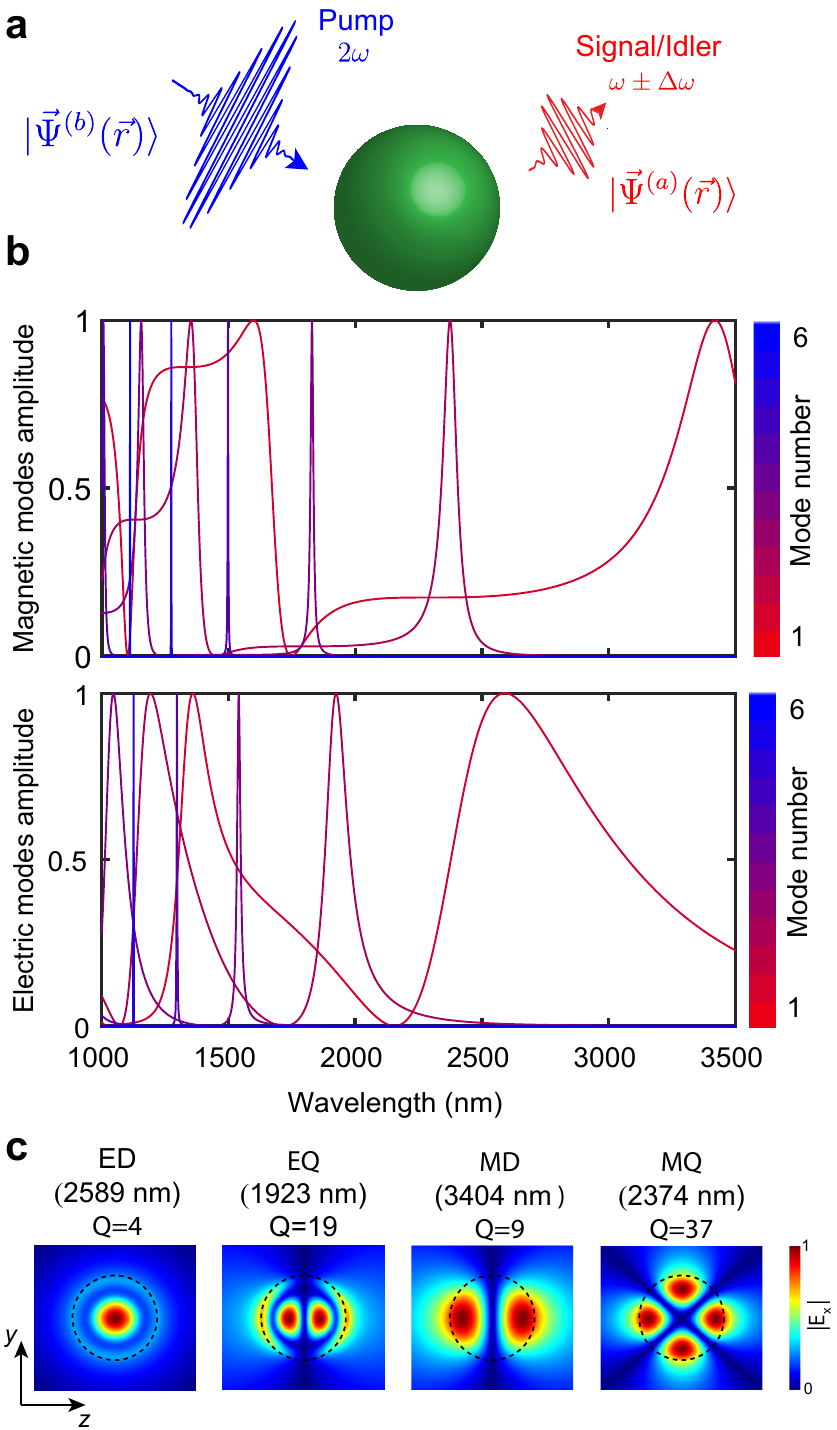}

\end{tabular}
\caption{{\bf OPO in a nanoscale dielectric sphere}. {\bf a,} The resonator is composed of a AlGaAs spherical particle with a radius of 500 nm. A pump with a wavelength around the particle size can excite the multipolar modes of the particle.  {\bf b,} Normalized scattering amplitude of the electric and magnetic modes. It is seen that for a broad portion of the spectrum, the particle supports multiple modes which spatially and spectrally overlap. {\bf c,} The normalized $x$ component of the electric field for the first two electric and the first two magnetic eigenmodes. If the pump is in the sub-wavelength-regime (pump wavelength is $>1500~$~nm), the OPO signal can be a superposition the lower order modes: electric dipole (ED), electric quadrupole (EQ), magnetic dipole (MD), and magnetic quadrupole (MQ). For the excitations at shorter wavelengths, higher order modes come into play as well. The resonant wavelength and the Q factor of the higher order modes are reported in the Supplementary Material.}
\label{fig:NanoOPO_Scattering}
\end{figure}

The wave equation for each of the signal modes is simplified to (see the Supplementary Material):
\begin{align}
\label{eq:ND_signal_multimode_main}
 \frac{d}{dt}a_l^{(s)} =\left(i\delta\omega_l^{(a)}-\frac{\alpha_l^{(a)}}{2}\right)a_l^{(s)}+ib\sum_k \eta_{lk}a^{(i)*}_k,
\end{align}
where $a^{(s)}$, $a^{(i)}$ and $b$ represent signal, idler, and pump envelope, respectively. $i\delta\omega_l^{(a)}$ and $\alpha_l^{(a)}$ are the detuning and the decay rate for the signal/idler modes, respectively, and $\eta_{lk}$ is the nonlinear coupling between the l\textsuperscript{th} mode and the k\textsuperscript{th} mode as:
\begin{align}
\label{nonlinear_coupling_main}
    \eta_{lk}=\omega\langle{\frac{\mathcal{E}_b\chi^{(2)}}{n(\omega)^2}\vec{\psi}^{(a)^*}_l(\vec{r})\vec{\Psi}^{(b)}(\vec{r})\vec{\psi}^{(a)^*}_k(\vec{r})}\rangle.
\end{align}
Note that the pump mode, $b(t)|{\vec{\Psi}^{(b)}(\vec{r})}\rangle$, is a superposition of modes at the pump wavelength which is dictated by the input excitation. However, the signal has to be expanded to the quasi-normal modes (See the Supplementary Material). 
Equation~\ref{eq:ND_signal_multimode_main} combined with a similar equation governing the idler dynamics can be written in a matrix form as:
\begin{align}
\label{eq:matrix_form_main}
    i\frac{d}{dt}\pmb{\mathcal{A}}(t)=\pmb{\mathcal{H}}(b)\pmb{\mathcal{A}}(t)
\end{align}
where $\pmb{\mathcal{A}}(t)=\left[a_1^{(s)},a_1^{(i)*},...,a_k^{(s)},a_k^{(i)*},... \right]^T$. The electric field can be expressed as a superposition of the eigenmodes as:
\begin{align}
    \vec{E}_\omega(\vec{r},t) = e^{-i\omega t}\sum_m \Bigl(&e^{-i\lambda_m t} {\sum_k a_{k,m}^{(s)}|{\vec{\psi}^{(a)}_k(\vec{r})}\rangle}\\ \nonumber
    + &e^{+i\lambda_m^* t} {\sum_k a_{k,m}^{(i)*}|{\vec{\psi}^{(a)}_k(\vec{r})}\rangle}\Bigr) + c.c.,
\end{align}
where $[\lambda_m]$ are the eigenvalues and $\vec{V}_m= [a_{k,m}^{(s,i)}]$ are the corresponding eigenvectors of the Hamiltonian ($\pmb{\mathcal{H}}$) which define the signal/idler supermodes. 
A supermode starts to oscillate when the imaginary part of the corresponding eigenvalue (${\rm Im}(\lambda_m)$) surpasses zero. The minimum pump power to reach this condition defines the oscillation threshold. The real part of the eigenvalues corresponds to the signal and idler frequency separation from the half-harmonic (${\rm Re}(\lambda_m)=\Delta\omega$; $\omega_{s,i}=\omega \pm \Delta\omega$). Hence, the eigenvalues for degenerate OPOs ($\omega_s=\omega_i=\omega_p/2$) are pure imaginary, and they are complex for non-degenerate cases. 

\begin{figure}
\centering
\begin{tabular}{cc}

\includegraphics[width=8cm]{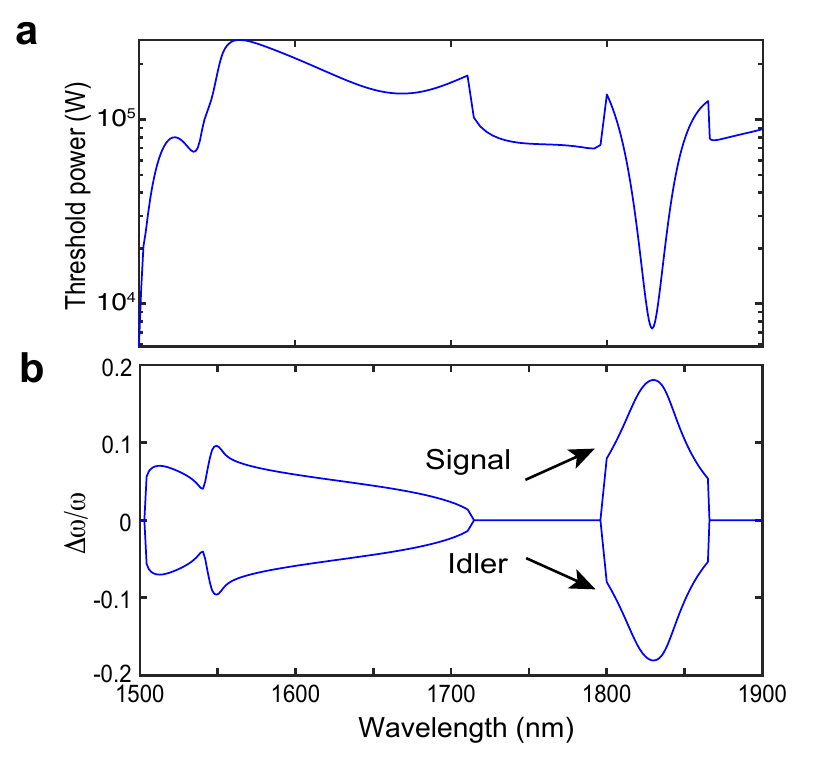}

\end{tabular}
\caption{{\bf Sub-wavelength OPO in a dielectric sphere}. The structure is the same as that shown in Fig.~\ref{fig:NanoOPO_Scattering}. {\bf a,} The oscillation threshold as a function of the pump wavelength. The input is a plane wave which excites multiple modes of the resonator at the pump wavelength. {\bf b,} The signal and idler frequency separation ($\pm\Delta\omega$) from the half-harmonic frequency ($\omega)$ at the threshold as a function of the pump wavelength.
The nonlinear interaction between the modes can reduce the threshold significantly. It can also cause a phase transition from degenerate ($\Delta\omega=0$) to non-degenerate ($\Delta\omega \ne 0$) which results in a sudden change in the oscillation threshold.}
\label{fig:NanoOPO_Subwavelength_Sweep}
\end{figure}

\begin{figure*}
\centering
\begin{tabular}{cc}

\includegraphics[width=16cm]{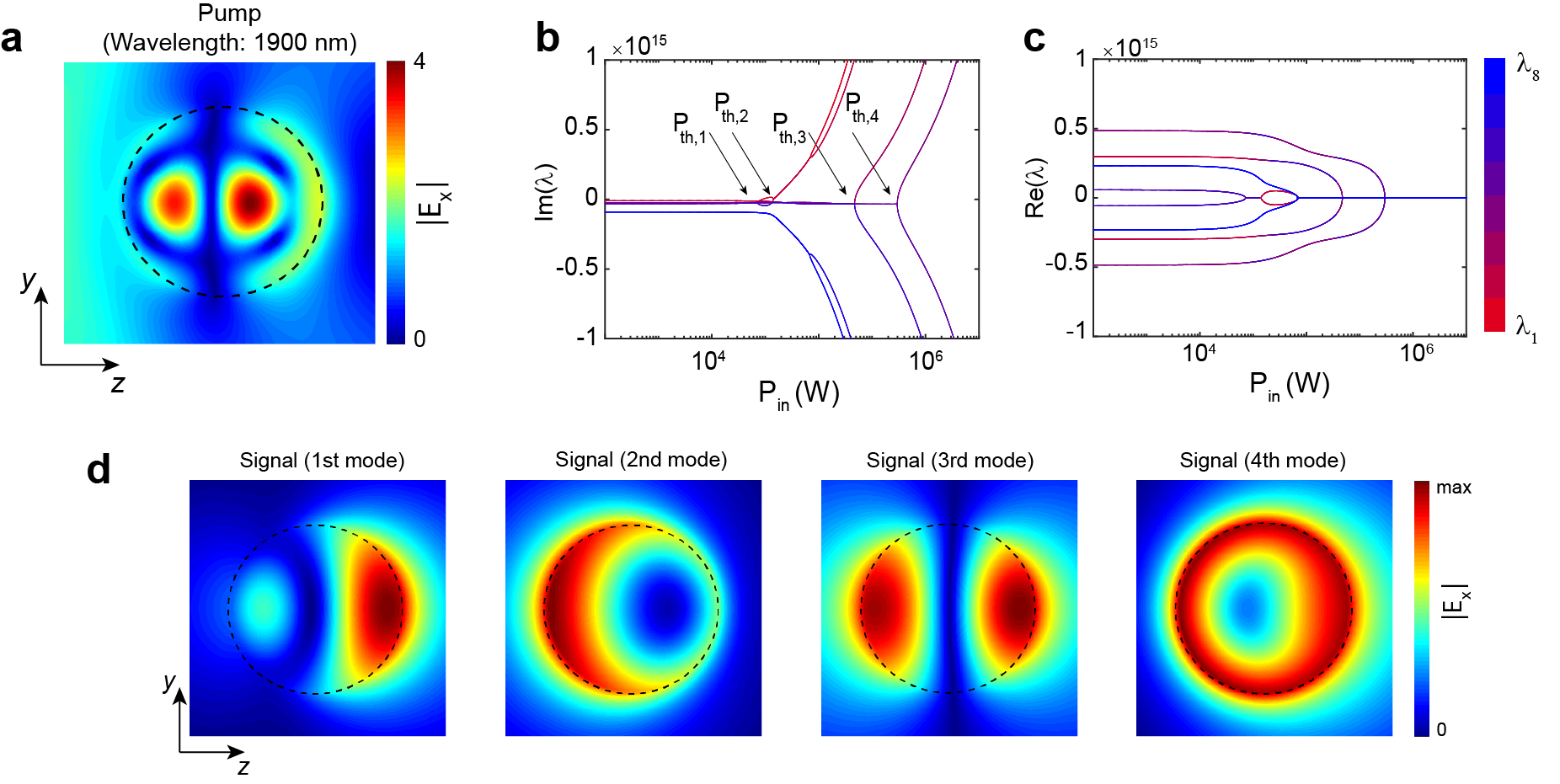}

\end{tabular}
\caption{{\bf Modes and eigenvalues in a sub-wavelength OPO in a dielectric sphere}. {\bf a,} The electric field profile at pump wavelength (1900~nm) normalized to the amplitude of the plane-wave excitation. {\bf b,c,} The imaginary (b) and the real (c) part of the eigenvalues as a function of the pump power. The eigenvalues are sorted based on their real part values. The imaginary part and the real parts correspond to the parametric gain and detuning from the half-harmonic ($\Delta\omega$), respectively.  When the real part of the eigenvalues becomes positive, the parametric gain overcomes the loss. Hence, the down-converted signal can surpass the oscillation threshold. {\bf d,} The electric field profile of the signal supermodes. It is seen that even though the detuning for MQ mode at half-harmonic is significantly larger compared to ED and MD modes, the contribution of MQ mode on the first signal supermode is more evident. This is due to the stronger overlap between the pump mode and the MQ mode. The strong nonlinear coupling between the signal eigenmodes helps to reduce the threshold 50 times compared to the case where we consider only one of the modes for the signal.} 
\label{fig:NanoOPO_Threshold_Pump}
\end{figure*}

\section{Results}

Our model is general and can be applied to a wide range of resonators. First, we apply our model to estimate the threshold in an AlGaAs sphere (Fig.~\ref{fig:NanoOPO_Scattering}a). The reason that we have chosen this simple structure is that the eigenmodes for this structure can be derived analytically and be expressed as multipolar resonances \cite{jahani2016all, kuznetsov2016optically}. Since the modes for a wide range of nano-structures, such as cylinders and cubes, can be expressed as multipolar resonances as well, our results can shed some light on the possibility of OPO in similar structures which are more amenable to fabrication on a chip \cite{baranov2017all, krasnok2018nonlinear, pertsch2020nonlinear, liu2016resonantly, carletti2018giant, marino2019spontaneous, timofeeva2018anapoles, koshelev2020subwavelength, gigli2020quasinormal}. Besides, AlGaAs is a low-loss high-index ($\varepsilon \approx 10$) material at optical frequencies with strong second-order nonlinearity ($\chi^{(2)}_{ijk}=200$ pm/V, $i \ne j \ne k$) \cite{gili2016monolithic}, and with appropriate orientation \cite{buckley2014multimode}, it has been recently explored for strong second-harmonic generation at nanoscale \cite{gili2016monolithic, smirnova2016multipolar,timofeeva2018anapoles,marino2019spontaneous,koshelev2020subwavelength}. Hence, it is an excellent candidate for demonstration of OPO at wavelength-scale with relatively low threshold. For a general case of dispersive or non-spherical three-dimensional resonators (Fig.~\ref{fig:NanoOPO_Schematic}a), we can use Lorentz reciprocity theory to find the quasi-normal modes of the resonator \cite{sauvan2013theory, raman2010photonic, lalanne2018light, yan2018rigorous}. The details are reported in the Supplementary Material. 

Figure~\ref{fig:NanoOPO_Scattering}b illustrates the normalized scattering coefficients for the first 6 electric and magnetic modes of a particle with a radius of 500 nm. If the particle is excited with a plane wave (or a Gaussian beam), several multipolar modes are excited. We first set the pump in the sub-wavelength regime (pump wavelength $>1500$~nm)  where lower order low-Q modes can be excited at the signal and idler frequencies. Then we discuss the behavior of the OPO in wavelength-scale (pump wavelength $\approx 1000$~nm) regime where higher order modes can also contribute.

If we operate in the sub-wavelength regime (i.e. the pump wavelength is larger than the particle size), only the first two electric and the first two magnetic modes can oscillate in the down-conversion process. Higher order modes can be neglected because of their large detuning ($\delta\omega_k \gg 1$).  The electric field profile of these four modes are illustrated in Fig. \ref{fig:NanoOPO_Scattering}c. The contribution of each mode in the OPO signal/idler supermode is dictated by the field overlap between the pump and the mode as well as the intermode nonlinear coupling as expressed in Eq.~\ref{nonlinear_coupling_main}, the Q factor, and the detuning from the half-harmonic frequency. Figure~\ref{fig:NanoOPO_Subwavelength_Sweep} displays the oscillation threshold as well as the spectral separation of signal and idler as a function of the pump wavelength. The dip in the threshold spectrum around 1830 nm is due to the enhancement of the pump field as a result of the excitation of the \nth{3} magnetic mode. Away from the center of the resonance, the input pump can still excite multiple lower order modes of the resonator. 

\begin{figure}
\centering
\begin{tabular}{cc}
\includegraphics[width=8.5cm]{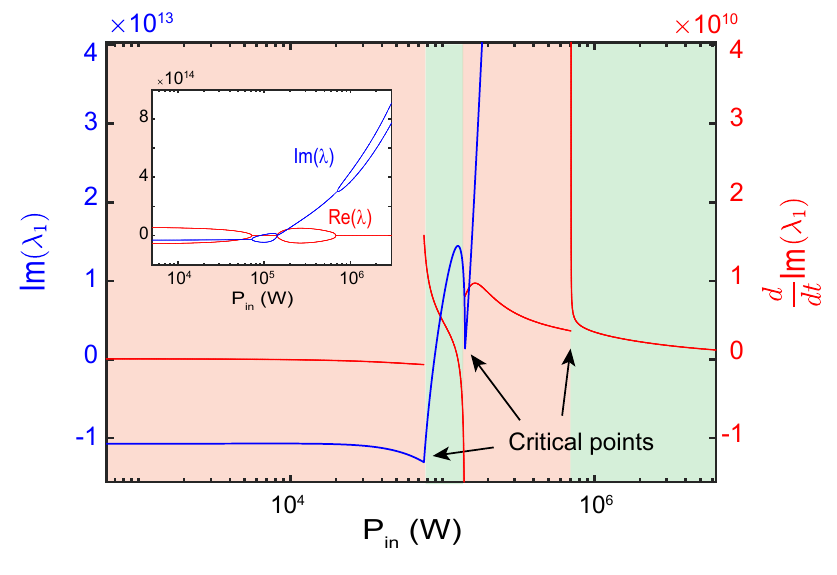}

\end{tabular}
\caption{{\bf Phase transition in wavelength-scale OPOs}. The imaginary part of the largest eigenvalue (blue) and its derivative for the structure shown in Fig.~\ref{fig:NanoOPO_Threshold_Pump}. The inset shows both the real and imaginary parts of the two eigenvalues with the smallest real part, which correspond to $\lambda_1$ and $\lambda_2$ for most of the input powers.
When there is a transition from degenerate (shaded as green) to non-degenerate (shaded as red), there is an abrupt change in the parametric gain at the critical points. The discontinuities in the derivative of the parametric gain corresponds to phase transitions in OPO. When two eigenvalues coalesce at a critical point, the derivative of the parametric gain diverges.} 
\label{fig:NanoOPO_Phase_transition}
\end{figure}

\begin{figure*}
\centering
\begin{tabular}{cc}

\includegraphics[width=17cm]{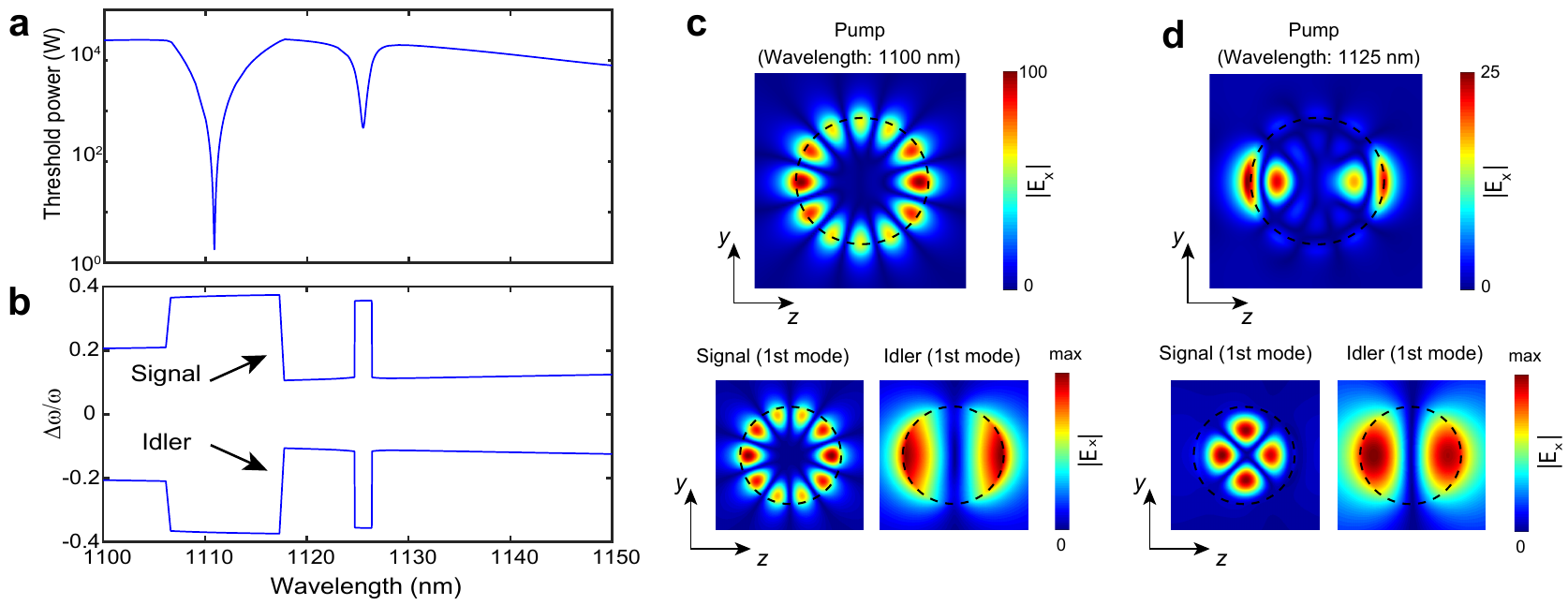}

\end{tabular}
\caption{{\bf Wavelength-scale OPO }. {\bf a,} The OPO threshold as a function of the pump wavelength. The dips in threshold around 1110 nm and 1125 nm correspond to the \nth{6} magnetic mode and the \nth{5} electric mode, respectively. The Q factor for these modes are 10$^4$ and 2500, respectively. The first 4 electric modes and 5 magnetic modes are considered as the eigenmodes for signal and idler modes. {\bf b,} Normalized detuning of the signal and idler from the half-harmonic at the threshold. Spectral phase transition can be observed leading to a sudden change in $\Delta\omega$ and a discontinuity in the derivative of the threshold. {\bf c,} The electric field distribution for the pump and the first signal/idler supermodes when the pump is at $1110$~nm. {\bf d,} The electric field distribution for the pump and the first signal/idler supermodes when the pump is at $1125$~nm.
}
\label{fig:NanoOPO_Threshold_Sweep}
\end{figure*}

If we ignore the intermode coupling and we assume that only one of the eigenmodes can oscillate, the OPO threshold would be considerably higher.  For instance, if the pump is at the center of the \nth{3} magnetic resonance, the minimum threshold for the single mode OPO is around 0.27 MW which is 36 times higher than the threshold shown in Fig.~\ref{fig:NanoOPO_Subwavelength_Sweep}a in which multi-mode interactions are taken into account (see the Supplementary Material for the threshold of all modes and coupling coefficients). In travelling-wave multi-mode OPOs, it is understood that, in the best case scenario, the threshold is of the order of the single-mode threshold divided by the number of modes \cite{de2006multimode}. The reason is that the modes in travelling wave resonators have the same nature. Thus, the maximum overlap is achieved if all the modes have the same mode profile \cite{de2006multimode, alves2018conditions}. However, in wavelength-scale OPOs, each of the multipolar modes have different spatial distribution, and their overlap through the pump field can potentially lead to a strong coupling even higher than the self coupling (the diagonal terms of $\eta_{lk}$). 

As seen in Fig.~\ref{fig:NanoOPO_Subwavelength_Sweep}, when the OPO goes through a transition from non-degenerate to degenerate oscillation, there is a sudden drop in the threshold. This corresponds to a phase transition from disordered to ordered phases which we have recently demonstrated in travelling wave OPOs \cite{roy2020spectral}. To understand the phase transition in wavelength-scale OPOs, we need to look at the eigenvalues and the eigenvectors of these resonators. For instance, we focus on a degenerate case with a pump excitation at 1900 nm (Fig.~\ref{fig:NanoOPO_Threshold_Pump}a). 
Figures~\ref{fig:NanoOPO_Threshold_Pump}b and\ref{fig:NanoOPO_Threshold_Pump}c display the real and imaginary parts of the eigenvalues as a function of input power, respectively. Since four modes are involved at signal and idler frequencies, there are eight eigenvalues and eight corresponding supermodes. The OPO threshold for each supermode is defined when the imaginary part of the eigenvalue passes zero (Fig.~\ref{fig:NanoOPO_Threshold_Pump}b).

At low input power levels, there is a weak coupling between the eigenmodes as seen in Eq.~\ref{eq:ND_signal_multimode_main}. Hence, each supermode is dominated by a single eigenmode (see the Supplementary Material for the eigenvectors). However, when the input power increases, the modes start to interact due to the nonlinear coupling through the pump. As a result, the supermodes near and above the threshold are a superposition of all eigenmodes. The electric field distribution of the four oscillating supermodes at the thresholds are shown in Fig.~\ref{fig:NanoOPO_Threshold_Pump}d. 

Moreover, due to the detuning of the center of resonance of the eigenmodes from the half-harmonic, the signal/idler supermodes for all eigenvalues are non-degenerate at low input power levels (${\rm Re}(\lambda_m) \ne 0$)  (Fig.~\ref{fig:NanoOPO_Threshold_Pump}c).  An increase in the input power enhances the intermode coupling which can change the signal and idler spectral separation. This can lead to a phase transition from non-degenerate to degenerate and vice versa. Especially at very high powers, the nonlinear coupling dominates over the detuning (Eq.~\ref{eq:ND_signal_multimode_main}), and as a result, all the modes are synchronized at the half-harmonic frequency (Fig.~\ref{fig:NanoOPO_Threshold_Pump}c).  

\begin{figure}
\centering
\begin{tabular}{cc}

\includegraphics[width=8.5cm]{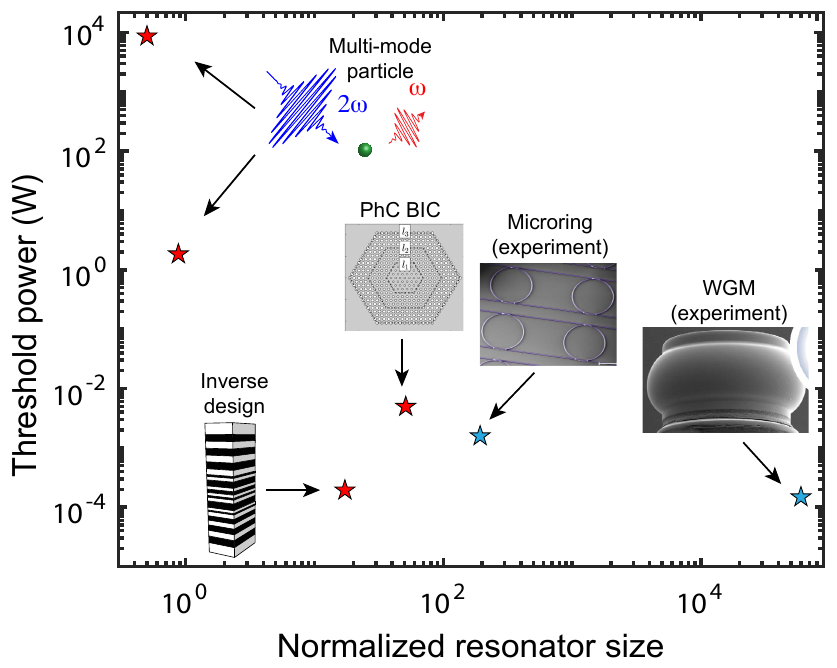}

\end{tabular}
\caption{{\bf Estimation of OPO threshold in various platforms}. The estimation of the threshold in a single-mode photonic crystal \cite{minkov2019doubly} and inversely designed cavities \cite{lin2016cavity} are based on the reported value for the SHG efficiency. The resonator sizes are normalized to the pump wavelength. As a reference, we have included OPOs demonstrated experimentally based on microring \cite{bruch2019chip} and whispering-gallery mode (WGM) \cite{werner2012blue} resonators.} 
\label{fig:NanoOPO_Comparison}
\end{figure}

The phase transition in the largest eigenvalue is illustrated in Fig.~\ref{fig:NanoOPO_Phase_transition}. This phase transition is accompanied by an abrupt change in the slope of the parametric gain \cite{roy2020spectral, Arnardottir2020PRL} which can be utilized for enhanced sensing and computing \cite{arute2019quantum, yang2020single}. A phase transition can happen due to either the competition between eigenvalues to achieve the highest gain or the coalescence of two eigenvalues.  
If a critical point is a coalescence of two eigenvalues, the eigenvectors coalesce as well at the critical point (see the Supplementary Material), which is a signature of exceptional points in non-Hermitian systems \cite{roy2020non, miri2019exceptional, parto2020non, lumer2013nonlinearly}.
We have recently shown first-order phase transition in coupled OPOs \cite{roy2020spectral}. However, the phase transition proposed here is observed in a single wavelength-scale OPO due to the strong nonlinear coupling between the multiple modes of the resonator.

To improve the performance of OPOs, it is desired to reduce the oscillation threshold further. The OPO threshold is inversely proportional to the Q factor of the pump mode if only one mode exists at the pump frequency (see the Supplementary Material). Hence, it is expected to reduce the threshold further by exciting the higher order modes as the higher order multipolar modes have even higher Q factor. Figure~\ref{fig:NanoOPO_Threshold_Sweep}a shows the OPO threshold for the first oscillating supermode as a function of the pump wavelength around the \nth{6} magnetic mode at 1110~nm with a Q factor of $10^4$ and the \nth{5} electric mode at 1125~nm with a Q factor of 2500. The seperation of the signal/idler frequencies from the half harmonic is shown in Fig.~\ref{fig:NanoOPO_Threshold_Sweep}b. For the signal and idler we have considered all the modes with a resonant wavelength longer than the pump wavelength (the first 4 electric and the first 5 magnetic modes).
The electric field distribution for the pump as well as the first signal/idler supermodes for the magnetic and electric mode excitations are shown in Figs.~\ref{fig:NanoOPO_Threshold_Sweep}c and \ref{fig:NanoOPO_Threshold_Sweep}d, respectively. The threshold at the center of the resonance of the \nth{6} magnetic mode and the \nth{5} electric mode can reach down to 2 W and 460 W, respectively.
Due to the large signal and idler separation, the parametric gain is low. However, at the input power of 43 W around the \nth{6} magnetic mode and the input power of 1900 W around the \nth{5} electric mode, the OPO experiences a phase transition into degenerate regime and the parametric gain is dramatically enhanced (see the Supplementary Material). Note that for the \nth{5} electric mode, even though the Q is large and high Q modes can also be excited at the signal wavelength, the threshold is not significantly different from the sub-wavelength regime shown in Fig.~\ref{fig:NanoOPO_Subwavelength_Sweep}. This is because of a weaker field overlap between the pump and signal modes in the absence of phase matching in larger resonators.
It is seen in Fig.~\ref{fig:NanoOPO_Threshold_Sweep}b that because of the competition between different eigenvalues, a phase transition can happen in non-degenerate regime with a sudden change in the signal/idler spectral separation resulting in a discontinuous change in the derivative of the OPO threshold as shown in Fig.~\ref{fig:NanoOPO_Threshold_Sweep}a. 

The approach that we have used to estimate the threshold can also be applied to estimate the second-harmonic generation in multi-mode wavelength-scale resonators (see the Supplementary Material for more details). 
Specifically, in the simple case that the pump and signal are single mode, we arrive at a simple relationship between the OPO threshold and the second-harmonic generation efficiency ($\epsilon_{\rm SHG}=P_{\rm out}/P_{\rm in}^2$, where $P_{\rm in}$ and $P_{\rm out}$ are the input power at the fundamental harmonic and output power at the second-harmonic, respectively) which can be expressed as:
\begin{align}\label{threshold_approx_main}
    P_{\rm th}=\frac{4{\alpha^{(a)}} ^2}{{\alpha^{(b)}}^2\epsilon_{\rm SHG}} \left( \frac{\frac{{\alpha^{(b)}}^2}{4}+\delta{\omega^{(b)}}^2}{\frac{{\alpha^{(a)}}^2}{4}+\delta{\omega^{(a)}}^2} \right) \approx \frac{4}{\epsilon_{\rm SHG}}.
\end{align}

As there is no threshold for SHG process and the conventional detectors are more sensitive at shorter wavelengths \cite{hadfield2009single}, it is usually easier to simulate or measure the SHG process. This allows us to estimate the OPO threshold in some structures which have already been proposed for SHG. Figure~\ref{fig:NanoOPO_Comparison} displays few examples and the estimated threshold in these structures. The low threshold in the inversely designed structure \cite{lin2016cavity} (compared to the photonic crystal structure which has a higher Q factor \cite{minkov2019doubly}) shows the importance of the field overlap to achieve a strong nonlinear response. Note that the thresholds reported in Fig.~\ref{fig:NanoOPO_Comparison} is for a continuous wave sources. 

Since the round-trip time in wavelength-scale OPOs is only few femto-seconds and the Q factor compared to micro-resonators is relatively low, the input pump can be compressed in time into a short pulse. This can lead to average-power thresholds of few tens of milliwatts (with a pulse repetition rate of 100 MHz) even for subwavelength OPOs, which is in the order of the threshold for free-space pulsed OPOs \cite{marandi2012coherence, muraviev2018massively}. Hence, the oscillation can happen before the onset of the material damage threshold.
The field overlap can be further enhanced by Mie resonance engineering \cite{bahng2020}, inverse design \cite{molesky2018inverse}, using hybrid plasmonic structures \cite{nielsen2017giant}, or controlling  evanescent waves \cite{jahani2018controlling}. This can potentially help to achieve sub-milliwatt oscillation threshold in subwavelength and wavelength-scale resonators.

\section{Conclusion}

In conclusion, we proposed a general theory to estimate the oscillation threshold in wavelength-scale OPOs and the nonlinear mixing behavior of modes above the threshold. We showed that the nonlinear interactions in multi-mode wavelength-scale resonators can be different from their large-scale counterparts and the threshold can be considerably reduced as a result of multi-mode interactions in these resonators. We demonstrated a phase transition in these resonators due to the nonlinear interactions between multiple modes. We have shown that although the phase matching is not required in this regime, the field overlap between modes can play a crucial role in reducing the threshold.
Our formalism is general and can predict the behavior of OPO above the threshold if the pump depletion is also taken into account. It can also be applied to $\chi^{(3)}$ cavities. Our approach can enable design of a new class of nonlinear integrated photonic systems.  
\\

{\bf \large Acknowledgment.} We thank Philippe Lalanne for valuable comments. The authors gratefully acknowledge support from ARO Grant No. W911NF-18-1-0285 and NSF Grant No. 1846273 and 1918549. The authors wish to thank NTT Research for their financial and technical support.
\\

{\bf \large Disclosures.} The authors declare no conflicts of interest.


\cleardoublepage
\renewcommand{\thefigure}{S\arabic{figure}}
\renewcommand{\thesection}{S.\arabic{section}}
\renewcommand{\theequation}{S\arabic{equation}}

\setcounter{section}{0}
\setcounter{figure}{0}
\setcounter{equation}{0}

\begin{widetext}
\begin{center}

\begingroup
    \fontsize{14pt}{12pt}\selectfont
    \bf{Wavelength-scale Optical Parametric Oscillators: Supplementary Material}
    
\endgroup
\bigskip
Saman Jahani, Arkadev Roy, and Alireza Marandi\\
\bigskip
Department of Electrical Engineering, 
California Institute of Technology, Pasadena, CA 91125, USA.
\\

\end{center}
{\small In the Supplementary Material, we derive the equations for single-mode and multi-mode OPOs for both degenerate and non-degenerate cases. We derive the second-harmonic generation (SHG) efficiency and establish a connection between the SHG efficiency and the threshold in degenerate OPOs for single mode cases. We discuss the quasi-normal modes for dispersive and non-spherical cases and the role of low-Q background modes on the performance of arbitrarily-shaped OPOs. We provide more details on the parameters, eigenvalues and eigenvectors of the results displayed in the main text.}

\end{widetext}

\section{Wave equations}

The Helmholtz wave equation in presence of nonlinear polarizability can be written as:

\begin{align}\label{wave_eqn}
    \nabla^2 \vec{E} &=\mu_0\frac{\partial}{\partial t}\left( \frac{\partial \vec{D}}{\partial t} +\sigma\vec{E}\right)\\ \nonumber
    &=\mu_0\varepsilon_0\varepsilon\frac{\partial^2 \vec{E}}{\partial t^2}+\mu_0\sigma\frac{\partial \vec{E}}{\partial t}+\mu_0\frac{\partial^2 \vec{P}_{\rm NL}}{\partial t^2},
\end{align}
where $\varepsilon=n^2$ is the linear relative permittivity, $n$ is the refractive index, and ${P}_{\rm NL}$ is the nonlinear polarization. To describe nonlinear dynamics in wavelength-scale cavities, we write the electric field as a superposition of the cavity eigenmodes. Instead of the conventional form of spatial SVEA in which the envelope evolves as the wave propagates through the nonlinear medium, we assume that the envelope is stationary in space but slowly evolves in time:
\begin{align}\label{E_expansion}
    \vec{E}(\vec{r},t) = \mathcal{E}_a &\sum_k a_k(t)e^{-i(\omega-i\frac{\alpha_k}{2})t}|{\vec{\psi}_k(\vec{r})}\rangle+c.c.,\\ \nonumber
    \vec{P}_{\rm NL}(\vec{r},t) = &\sum_k  \vec{P}_k(\vec{r},t) e^{-i(\omega-i\frac{\alpha_k}{2})t}+c.c.,
\end{align}
where $\mathcal{E}_a$ is the normalization constant such that $|a_k|^2$ is the energy stored in the k$^{th}$ mode of the cavity, and for a homogeneous resonator, it is $\mathcal{E}_a=\sqrt{2/\varepsilon_0 n(\omega)^2}$, $\vec{P}_k$ is the nonlinear polarization that we explain later, $|{\vec{\psi}_k(\vec{r})}\rangle$ is the cavity eigenmode normalized such that $\langle{\vec{\psi}_m(\vec{r})\vec{\psi}_k(\vec{r})}\rangle=\delta_{mk}$ ($\delta_{mk}$ is the Kronecker delta), $\omega$ is the angular frequency of the signal, idler or pump, $\alpha_k=\omega_k/Q_k$ is the decay rate of the cavity mode, $\omega_k$ is the eigenfrequency of the $k$-th mode with a quality factor of $Q_k$. 

In the following, we first formulate the nonlinear dynamics for a single-mode OPO at degeneracy, and then we expand the formalism to a multi-mode cavity and non-degenerate case. 

By inserting Eq.~\ref{E_expansion} in to Eq.~\ref{wave_eqn}, considering the k$^{th}$ mode is the only mode at the operating frequency, we have:
\begin{align}
     &\{ \nabla^2 + \frac{\omega^2}{c^2}n^2-\frac{n^2}{c^2}\frac{\partial^2}{\partial t^2}+ \frac{2i\left(\omega-i\frac{\alpha_k}{2}\right)}{c^2}n^2\frac{\partial}{\partial t} \\ \nonumber
    &+\frac{i\alpha_k\omega-\alpha_k^2/4}{c^2}n^2+i\omega\mu_0\sigma+\mu_0\sigma\frac{\partial}{\partial t} \} \mathcal{E}_a a_k(t)|{\vec{\psi_k}(\vec{r})}\rangle\\ \nonumber
    &=-\mu_0\left(\omega-i\frac{\alpha_k}{2}\right)^2\vec{P}_k+2i\mu_0\left(\omega-i\frac{\alpha_k}{2}\right)\frac{\partial\vec{P}_k}{\partial t}+\mu_0\frac{\partial^2\vec{P}_k}{\partial t^2}.
\end{align}

Because of SVEA, $\omega \gg \alpha_k$, $\omega P_k\gg \frac{\partial P_k}{\partial t}$, and $\omega a_k\gg \frac{\partial a_k}{\partial t}$. Also, if we ignore the effect of the nonlinearity on the dispersion and if we assume that $\omega= \omega_k+\delta\omega_k$ where $\omega_k \gg \delta\omega_k$, we can assume $\left( \nabla^2+\frac{\omega_k^2}{c^2}n^2 \right) |{\vec{\psi_k}(\vec{r})}\rangle\approx 0$. With these approximations, the wave equation is simplified to:
\begin{align}
     &\{\frac{2i\omega n^2 }{c^2}\frac{\partial}{\partial t} +i\omega\mu_0\sigma a_k\\ \nonumber
    &+\frac{(2\delta\omega_k+i\alpha_k) \omega n^2 }{c^2}\}\mathcal{E}_a a_k(t)|{\vec{\psi}_k(\vec{r})}\rangle=-\mu_0\omega^2\vec{P}_k.
\end{align}
Dividing the both sides by ${2i\omega n^2 }/{c^2}$, we reach:
\begin{align}
   \{\frac{\partial}{\partial t}+\frac{\mu_0\sigma c^2}{2}-i\delta\omega_k+\frac{\alpha_k}{2}\} \mathcal{E}_a a_k(t)|{\vec{\psi}_k(\vec{r})}\rangle=\frac{i{\mu }_0\omega c^2}{2n^2}\vec{P}_k. 
\end{align}

Note that we have assumed a weak material dispersion to derive the above equation. For dispersive structures, the evolution of modes need more rigorous analysis \cite{yan2018rigorous}.
We first implement the nonlinear dynamics to estimate the threshold in single-mode OPOs. Then, we extend our model when the cavity has multiple modes at the signal wavelength. We also applies our model for second-harmonic generation, we show that if the second-harmonic signal is single-mode, we can estimate the threshold from SHG efficiency. This can be helpful to estimate the OPO threshold for the structures which have already been proposed for SHG.

\section{Half-harmonic generation}

By writing the nonlinear polarization, we can find the nonlinear dynamics for different nonlinear processes (e.g. second-harmonic generation and half-harmonic generation). Here, we first focus on the threshold for half-harmonic generation in degenerate OPOs. For simplicity, we ignore the ohmic loss of the modes.

The coupled nonlinear wave equation for signal and pump can be written as:
\begin{align}
    \label{ND_signal}
    &\sum_k \{\frac{\partial}{\partial t}-i\delta\omega^{(a)}_k+\frac{\alpha^{(a)}_k}{2}\}a_k(t)|{\vec{\psi}^{(a)}_k(\vec{r})}\rangle\\ \nonumber
     = &\sum_k \frac{i\omega}{2n(\omega)^2}\chi^{(2)}(2\omega,\omega,\omega) \mathcal{E}_b b(t)a^*_k(t)|{\vec{\Psi}^{(b)}(\vec{r})|\vec{\psi}^{(a)^*}_k(\vec{r}})\rangle,\\
    \label{ND_pump}
    &\{\frac{\partial}{\partial t}-i\delta\omega^{(b)}+\frac{\alpha^{(b)}}{2}\}b(t)|{\vec{\Psi}^{(b)}(\vec{r})}\rangle\\ \nonumber 
    = &\sum_k \frac{i\omega}{n(2\omega)^2}\chi^{(2)}(2\omega,\omega,\omega) \frac{\mathcal{E}_a^2}{\mathcal{E}_b} a^2_k(t)|{\vec{\psi}^{(a)^2}_k(\vec{r})}\rangle. 
\end{align}
We have defined the electric field for the signal at the fundamental harmonic as $\vec{E}_\omega= \mathcal{E}_a\sum a_k(t)e^{-i(\omega-i\frac{1}{2}\alpha^{(a)}_k)t}|{\vec{\psi}^{(a)}_k(\vec{r})}\rangle$, where $|{\vec{\psi}^{(a)}_k(\vec{r})}\rangle$ are the eigenmodes of the cavity at $\omega=\omega_k$ with decay constant of $\alpha_k^{(a)}$. The electric field for the pump at second-harmonic is defined as $\vec{E}_{2\omega}=\mathcal{E}_be^{-i(2\omega-\frac{1}{2}\alpha^{(b)})t}b(t)|{\vec{\Psi}^{(b)}(\vec{r})}\rangle$, where $|{\vec{\Psi}^{(b)}(\vec{r})}\rangle$ is the spatial mode profile of the pump normalized such that $\langle{\vec{\Psi}^{(b)}(\vec{r})\vec{\Psi}^{(b)}(\vec{r})}\rangle=1$ but, as we explain later, it does not have to be the eigenmode of the cavity and it can be an embedded eigenmode of the cavity, such as Fano, anapole, or bound-state in the continuum modes, $b(t)$ is the envelope of the pump such that $|b|^2$ is the pump power, and $\alpha^{(b)}$ is the decay rate for the pump mode.

\subsection{Single-mode cavity}
If $|{\vec{\psi}^{(a)}_k(\vec{r})}\rangle$ is the only mode of the cavity at the operating frequency, by multiplying the both sides of Eqs.~\ref{ND_signal} and \ref{ND_pump} by $\langle{\vec{\psi}^{(a)}_k(\vec{r})}|$ and $\langle{\vec{\Psi}^{(b)}(\vec{r})}|$, respectively, and calculating the inner product, the coupled equations are simplified to:
\begin{align}
\label{ND_signal_simplified}
    \frac{d}{dt}a_k &=\left(i\delta\omega_k^{(a)}-\frac{\alpha_k^{(a)}}{2}\right)a_k+i\eta_{kk}ba^*_k,\\ 
    \label{ND_pump_simplified}
    \frac{d}{dt}b &=\left(i\delta\omega^{(b)}-\frac{\alpha^{(b)}}{2}\right)(b-b_0)+i2\eta^*_{kk}a_k^2,
\end{align}
where $b_0$ is the pump amplitude in the absence of the nonlinearity and $\eta_{lk}$ is the effective nonlinear coupling defined as:
\begin{align}
\label{nonlinear_coupling}
    \eta_{lk}=\omega\langle{\frac{\mathcal{E}_b\chi^{(2)}}{n(\omega)^2}\vec{\psi}^{(a)^*}_l(\vec{r})\vec{\Psi}^{(b)}(\vec{r})\vec{\psi}^{(a)^*}_k(\vec{r})}\rangle.
\end{align}

Near the OPO threshold, we can assume that the pump is not depleted ($b=b_0$). Above threshold, Eqs.~\ref{ND_signal_simplified} and \ref{ND_pump_simplified} must be solved simultaneously. The steady-state amplitude of the signal is the solution of Eq.~\ref{ND_signal_simplified} when $da_k/dt=0$. There are two solutions: one of them is the trivial solution, $a_k=0$, which represents the OPO below the threshold; the nontrivial solution which represents the OPO at threshold. This requires that the amplitude and phase of the pump satisfy these conditions:
\begin{align}
    |\eta_{kk}b_0|\sin{(\phi_b-2\phi_k)} &=\frac{\alpha_k^{(a)}}{2}, \\ \nonumber
    |\eta_{kk}b_0|\cos{(\phi_b-2\phi_k)} &=-\delta\omega_k^{(a)},
\end{align}
where $\phi_k$ and $\phi_b$ are the phase of the signal mode and the pump mode, respectively. As far as the threshold power is concerned, the above equation can be written in a more compact form \cite{haus1984waves,rodriguez2007chi}:
\begin{align}\label{threshold}
    |b_0|^2=\frac{1}{|\eta_{kk}|^2}\left(\frac{{\alpha_k^{(a)}}^2}{4}+{\delta\omega_k^{(a)}}^2\right).
\end{align}
If there is only one coupling channel between the input source and the cavity mode at the pump frequency, in the weak coupling regime ($Q_k \gg 1$), the coupling between the input source and the pump cavity mode in the steady-state can be written as \cite{rodriguez2007chi}:
\begin{align}
    |b_0|^2=\frac{\alpha^{(b)}}{\frac{{\alpha^{(b)}}^2}{4}+{\delta\omega^{(b)}}^2}P_{\rm in}.
\end{align}
Hence, the threshold for the input source to go above threshold is:
\begin{align}\label{threshold_Pin}
    P_{\rm th}=\frac{1}{\alpha^{(b)}|\eta_{kk}|^2}\left(\frac{{\alpha_k^{(a)}}^2}{4}+{\delta\omega_k^{(a)}}^2\right)\left(\frac{{\alpha^{(b)}}^2}{4}+{\delta\omega^{(b)}}^2\right).
\end{align}
If there are more than one coupling channel between the input and the cavity, such as the excitation from the free-space, Eq.~\ref{threshold_Pin} is not accurate, and the coupling between the input power and the pump mode amplitude, $b_0$, should be derived from the linear analysis of the cavity at the pump frequency.  

\begin{figure*}
\centering
\begin{tabular}{cc}
\includegraphics[width=18cm]{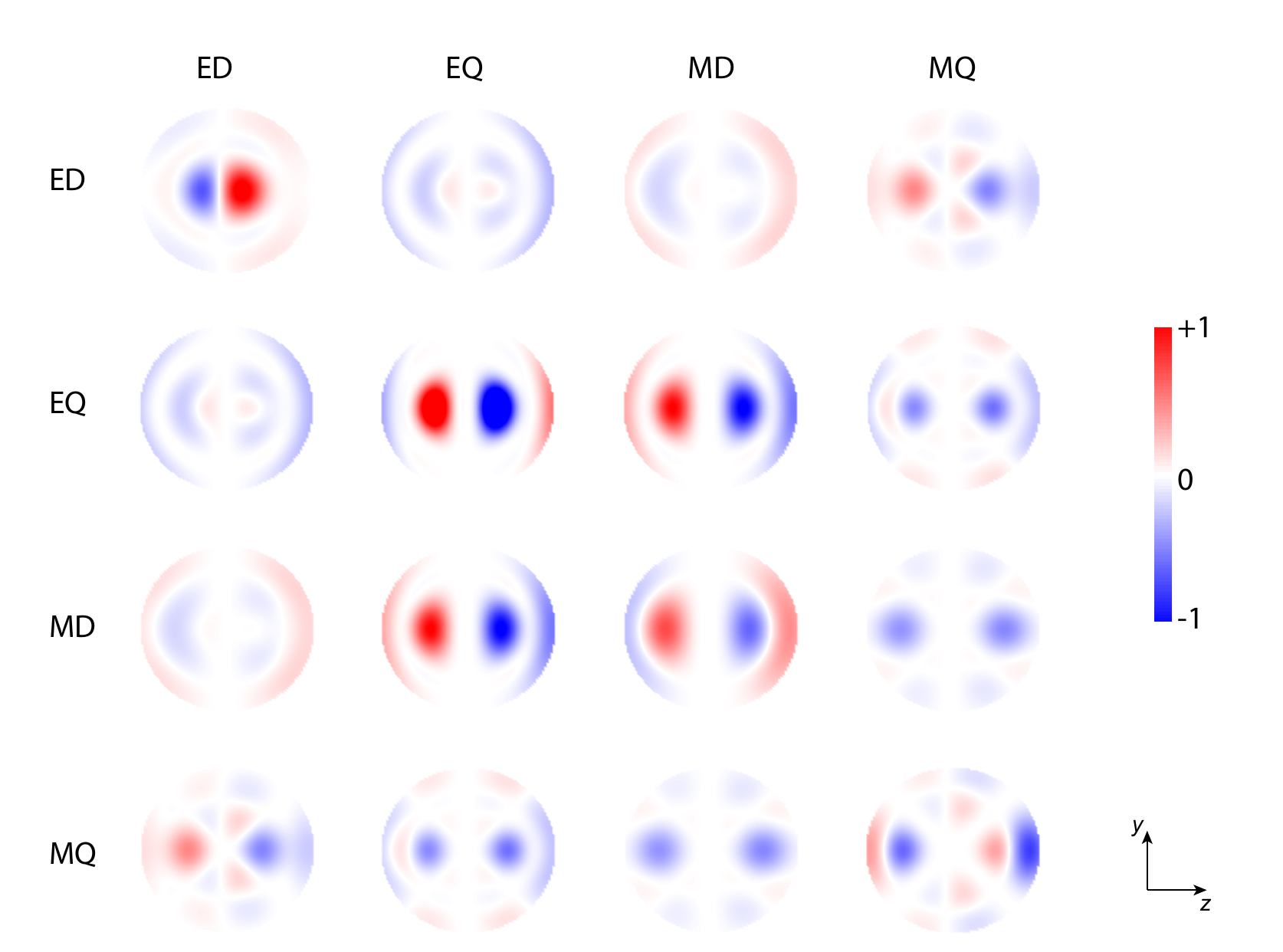}

\end{tabular}
\caption{{\bf Field overlap between the pump, signal, and idler}.
Normalized local field overlap, which is defined in Eq.~\ref{nonlinear_coupling}, between the pump (for the structure shown in Fig. 3 in the main text when the pump is resonating at the \nth{3} magnetic mode), signal (rows), and idler (columns) modes in the nonlinear region. The integration of the overlap leads to the nonlinear coupling matrix demonstrated in Eq.~\ref{eq:etta}. It is seen that even though the intensity of local overlap for some modes are high, the nonlinear coupling is small due to the weak overlap between the mode profiles. This means that increasing the field intensity locally (e.g. by increasing the Q factor) does not always lead to a stronger nonlinear response in wavelength-scale resonators.} 
\label{fig:NanoOPO_Overlap}
\end{figure*}

\begin{figure*}
\centering
\begin{tabular}{cc}

\includegraphics[width=17cm]{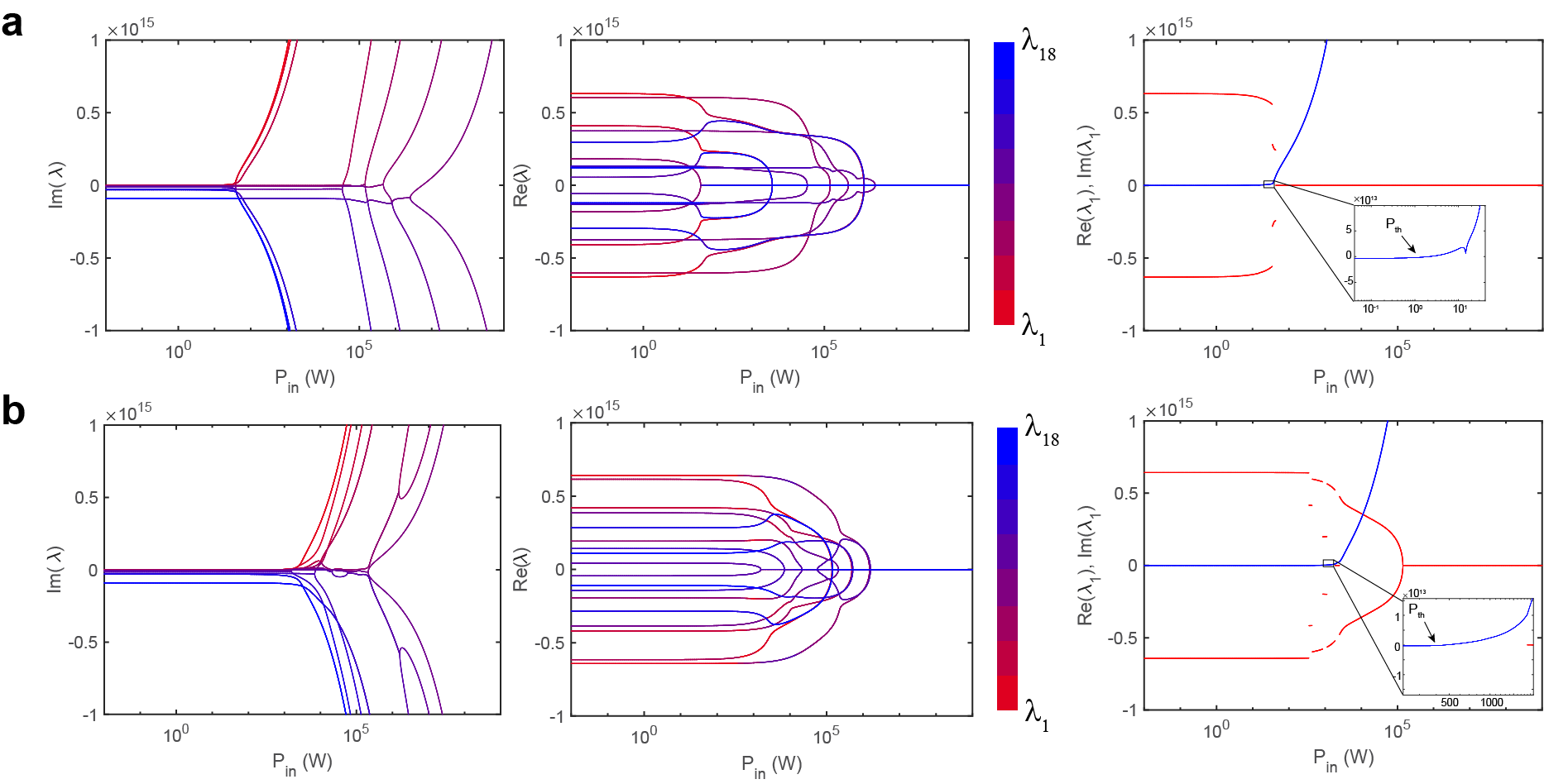}

\end{tabular}
\caption{{\bf Eigenvalues of the wavelength-scale OPOs}.
The structure is the same as that Fin. 6 in the main text. {\bf a,} the pump wavelength is at 1110 nm. {\bf b,} the pump wavelength is at 1125 nm. The OPO threshold is 2W and 467 W, respectively. However, the parametric gain is small because of the large signal/idler separation. As soon as the OPO goes through a phase transition into degenerate phase, the parametric gain increases rapidly. The threshold for degenerate OPO is 34 W and 1929 W, respectively.} 
\label{fig:NanoOPO_Eigenvalues_1110_1125}
\end{figure*}

\subsection{Multi-mode cavity}

For wavelength-scale cavities, the quality factor of the modes are usually low. Hence, at operating wavelength more than one can resonate. If the cavity is multi-mode at the operating wavelength, by multiplying the both sides of Eq.~\ref{ND_signal} by $\langle{\vec{\psi}^{(a)}_l(\vec{r})}$, the coupled equation is simplified to:

\begin{align}
\label{eq:D_signal_multimode}
    \frac{d}{dt}a_l &=\left(i\delta\omega_l^{(a)}-\frac{\alpha_l^{(a)}}{2}\right)a_l+ib\sum_k \eta_{lk}a^*_k.
\end{align}
The steady-state response of this equation can be written in a matrix form as:
\begin{align}
\label{eq:matrix_form}
    \pmb{\mathcal{H}}(b)\left[a_1,a_1^*,...,a_k,a_k^*,... \right]^T=0.
\end{align}
The OPO threshold is the minimum pump power for which the determinant of the matrix passes zero. Near the threshold, that is the only oscillating mode and the eigenvector correspond to that eigenvector describes the spatial distribution of the signal. The phase difference between each mode of the pulse and the pump is set automatically to achieve the minimum threshold. There is no closed form solution for the eigenvalue if the quality factors of the modes or the central frequencies of all modes are not the same. However, in the best case scenario where all the modes have similar nonlinear coupling coefficient and quality factor, the threshold is reduced by a factor which is the number of modes. 

As seen in Figs. 3 and 6 in the main text, the threshold for degenerate OPO is not always lower than the non-degenerate case. Hence, it is crucial to consider non-degenerate cases as well.

If signal and idler modes are non-degenerate, Eq.~\ref{eq:D_signal_multimode} is changed to:
\begin{align}
\label{eq:ND_signal_multimode}
 \frac{d}{dt}a_l^{(s)} =\left(i\delta\omega_l^{(a)}-\frac{\alpha_l^{(a)}}{2}\right)a_l^{(s)}+ib\sum_k \eta_{lk}a^{(i)*}_k,
\end{align}
where $a_l^{(s)}$ and $a_l^{(s)}$, represent the envelope of the $l^{th}$ signal and idler mode, respectively. In this case, the eigenvalues are not necessarily real, and the steady-state response can be oscillatory. As a result, the eigenvalue problem of Eq.~\ref{eq:matrix_form} is changed to:
\begin{align}
\label{eq:matrix_form_main}
    i\frac{d}{dt}\pmb{\mathcal{A}}(t)=\pmb{\mathcal{H}}(b)\pmb{\mathcal{A}}(t)
\end{align}
where $\pmb{\mathcal{A}}(t)=\left[a_1^{(s)},a_1^{(i)*},...,a_k^{(s)},a_k^{(i)*},... \right]^T$. The electric field for both degenerate and non-degenerate cases can be written as:
\begin{align}
    \vec{E}_\omega(\vec{r},t) = e^{-i\omega t}\sum_m \Bigl(&e^{-i\lambda_m t} {\sum_k a_{k,m}^{(s)}|{\vec{\psi}^{(a)}_k(\vec{r})}\rangle}\\ \nonumber
    + &e^{+i\lambda_m^* t} {\sum_k a_{k,m}^{(i)*}|{\vec{\psi}^{(a)}_k(\vec{r})}\rangle}\Bigr) + c.c.,
\end{align}
where $[\lambda_m]$ are the eigenvalues and $\vec{V}_m= [a_{k,m}^{(s,i)}]$ are the corresponding eigenvectors of the Hamiltonian ($\pmb{\mathcal{H}}$) which define the signal/idler supermodes. 

\begin{figure*}
\centering
\begin{tabular}{cc}

\includegraphics[width=17cm]{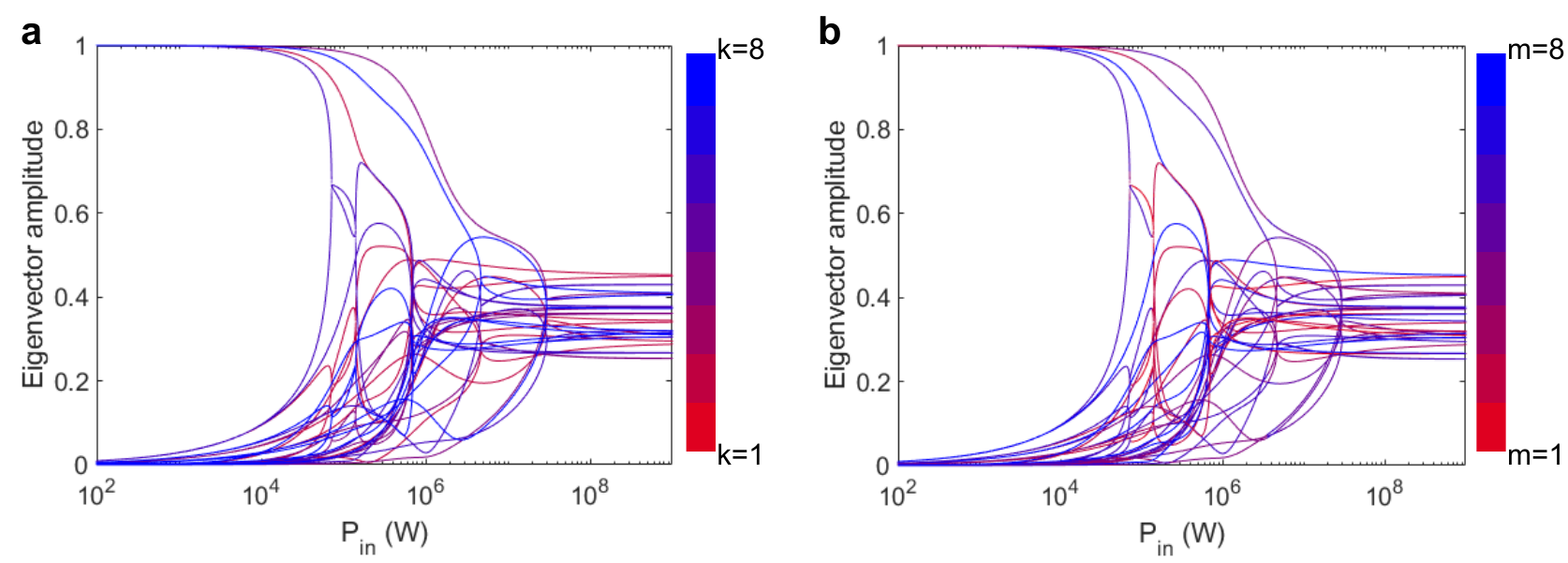}

\end{tabular}
\caption{{\bf Eigenvector amplitudes, $a_{k,m}^{(a)}$, for all eigenvalues}.
The structure is the same as that shown in Fig.~4 in the main text.
{\bf a} The color bar represents the mode number. {\bf b} The color bar represents the eigenvalue number.
It is seen that at low input power, each supermode corresponds to one eigenmode. However, as we approach the threshold  the nonlinear coupling increases the contribution of other eigenmodes for all signal supermode.} 
\label{fig:NanoOPO_Eigenvectors_all}
\end{figure*}

\section{Second-harmonic generation}

We can implement the same approach for calculating the SHG in cavities. However, for SHG, we have to expand the second-harmonic mode into the eigenmodes of the cavity while the pump input at fundamental harmonic can be an embedded mode of the cavity. If we ignore the back conversion, the nonlinear dynamic for SHG process can be written as:
\begin{align}
    \label{ND_SHG}
    &\sum_k\{\frac{\partial}{\partial t}-i\delta\omega^{(b_k)}+\frac{\alpha_k^{(b)}}{2}\}b_k(t)|{\vec{\psi}_k^{(b)}(\vec{r})}\rangle\\ \nonumber 
    &= \frac{i\omega}{n^2}\chi^{(2)}(2\omega,\omega,\omega) a^2(t)|{\vec{\Psi}^{(a)}(\vec{r})}\rangle^2. 
\end{align}
By multiplying the both sides by $\langle{}{\vec{\psi}_k^{(b)}}|$. Eq.~\ref{ND_SHG} is simplified to:
\begin{align}
    \label{ND_SHG_b}
    \frac{d}{dt}b_k &=\left(i\delta\omega_k^{(b)}-\frac{\alpha_k^{(b)}}{2}\right)b_k+i2\tilde{\eta}^*_ka^2,
\end{align}
where $\tilde{\eta}_k=\omega \langle{\mathcal{E}\frac{\chi^{(2)}}{n^2}\vec{\Psi}^{(a)^*}(\vec{r})^2\vec{\psi}^{(b)}_k(\vec{r})}\rangle$. If we assume that the pump is constant ($a(t)=a_0$), the steady-state second-harmonic generated power is:
\begin{align}\label{SHG_cavity_mode}
    |b_k|^2=\frac{4\tilde{\eta}_k^2}{\frac{{\alpha_k^{(b)}}^2}{4}+{\delta\omega_k^{(b)}}^2}|a_0|^4
\end{align}
If there is only one coupling channel between the input and the cavity mode at the fundamental frequency, the cavity mode amplitude can be written as the input power as:
\begin{align}\label{input_to_output}
    |a_0|^2&=\frac{\alpha^{(a)}}{\frac{{\alpha^{(a)}}^2}{4}+{\delta\omega^{(a)}}^2}P_{\rm in}, \\ \nonumber
    |b_k|^2&=\frac{\alpha_k^{(b)}}{\frac{{\alpha_k^{(b)}}^2}{4}+{\delta\omega_k^{(b)}}^2}P_{{\rm SHG},k}.
\end{align}
By inserting Eq.~\ref{input_to_output} in to Eq.~\ref{SHG_cavity_mode}, the second-harmonic power can be expressed as $P_{{\rm SHG}, k}=\epsilon_{{\rm SHG}, k}P_{\rm in}^2$, where $\epsilon_{{\rm SHG}}$ is the SHG efficiency in the unit of W$^{-1}$ written as:
\begin{align}\label{SHG efficiency}
    \epsilon_{{\rm SHG}, k}=\frac{4\tilde{\eta}_k^2{\alpha^{(a)}}^2}{\alpha_k^{(b)}\left(\frac{{\alpha^{(a)}}^2}{4}+{\delta\omega^{(a)}}^2\right)^2}.
\end{align}
If the cavity is single mode at both the fundamental and second harmonic, $\tilde{\eta}_k=\eta_{kk}$. This allows us to connect the SHG efficiency to the nonlinear coupling coefficient. Hence, by knowing the linear response of the cavity and SHG efficiency, we can derive the OPO threshold by inserting Eq.~\ref{SHG efficiency} into Eq.~\ref{threshold_Pin}:
\begin{align}\label{threshold_approx}
    P_{\rm th}=\frac{4{\alpha^{(a)}} ^2}{{\alpha^{(b)}}^2\epsilon_{\rm SHG}} \left( \frac{\frac{{\alpha^{(b)}}^2}{4}+\delta{\omega^{(b)}}^2}{\frac{{\alpha^{(a)}}^2}{4}+\delta{\omega^{(a)}}^2} \right) \approx \frac{4}{\epsilon_{\rm SHG}}.
\end{align}

\begin{figure*}
\centering
\begin{tabular}{cc}

\includegraphics[width=17cm]{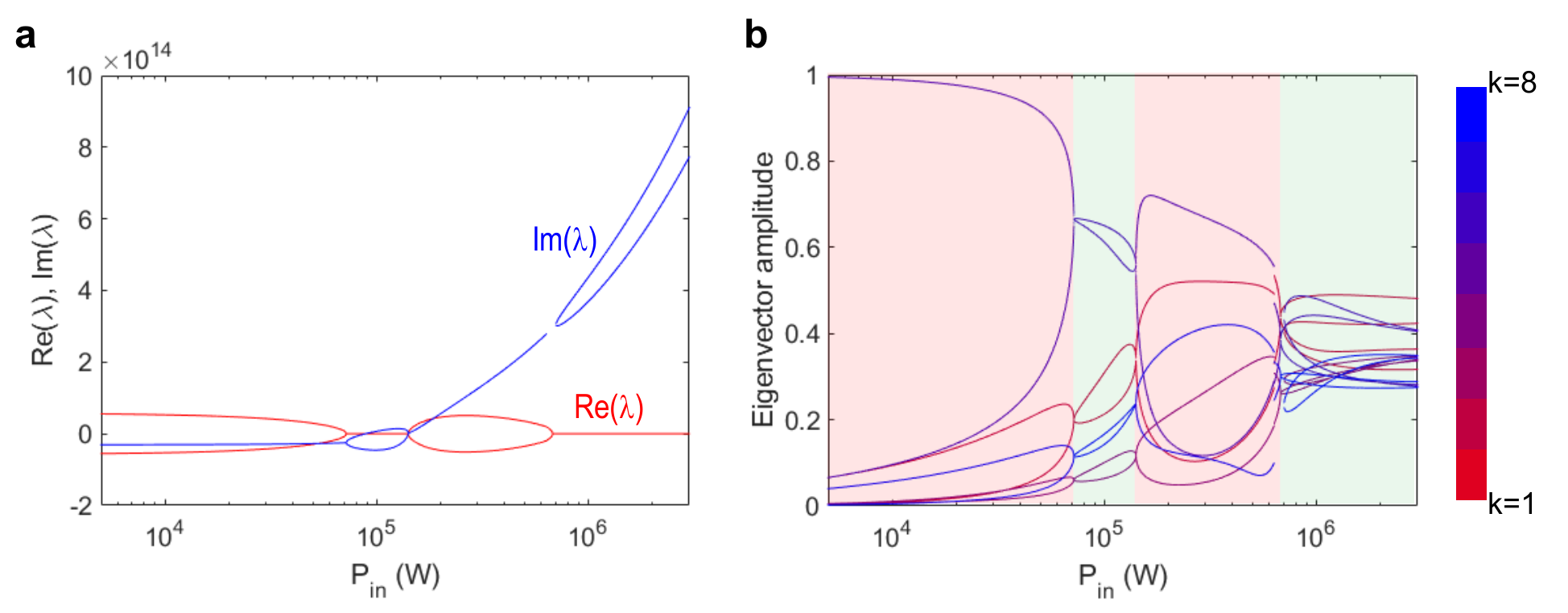}

\end{tabular}
\caption{{\bf Eigenvector amplitudes, $a_{k,m}^{(a)}$, for all eigenvalues}.
The structure is the same as that shown in Fig.~4 in the main text.
{\bf a} The real and the real parts of the two eigenvalues with the smallest real parts (as well as the largest imaginary parts for most of the input powers) {\bf b} The eigenvectors for the corresponding supermodes. The degenerate and non-degenerate regions are shaded as green and red, respectively. The eigenvalues as well as the eigenvectors of the two modes coalesce when there is a transition from degenerate to non-degenerate cases and vice versa.} 
\label{fig:NanoOPO_Exceptional_Points}
\end{figure*}

\section{OPO in spherical dielectric particle}
The nonlinear coupling matrix (Eq.~\ref{nonlinear_coupling}) for the particle shown in Fig. 3 in the main text when the pump is at the resonant frequency of the \nth{3} magnetic mode is calculated as:

\begin{align}
\label{eq:etta}
   |\eta_{lk}|=10^{4} \times
\begin{bmatrix} 
    0.5958 &   1.2898 &   1.1236 &   0.2458\\
    1.2898 &   0.0454 &   0.1493 &   2.4897\\
    1.1236 &   0.1493 &   0.4892 &   8.5686\\
    0.2458 &   2.4897 &   8.5686 &   0.7508\\
\end{bmatrix} 
\end{align}

The modes are ordered as: ED, EQ, MD, and MQ. The field overlap before the integration is shown in Fig.~\ref{fig:NanoOPO_Overlap}. It is seen that in the absence of the phase matching in wavelength-scale resonators, a strong local intensity of the overlap does not necessarily lead to a strong nonlinear coupling between two modes. This can also cause the off-diagonal terms to be stronger than the diagonal terms. If we ignore intermode coupling (off-diagonal terms), the threshold for these modes are: 3.99, 2783, 0.27, and 3.65 MW, respectively. However, due to the strong intermode coupling, which can be even stronger than the diagonal terms based on Eq.~\ref{eq:etta}, the threshold is reduced 36-fold as shown in Fig. 3 in the main text.

For the wavelength-scale OPO reported in Fig.~6 of the main text, there are 9 eigenmodes involved. The resonant wavelength of these modes are: 2589, 1923, 1541, 1297, 3404, 2374, 1829, 1498, and 1273 nm. The first 4 modes are the electric modes and the last 5 modes are the magnetic modes. They are sorted from the lowest order to the highest order. The Q factor of these modes are 4, 19, 100, 520, 9, 37, 141, 600, and 2500, respectively. The nonlinear coupling term for the pump excitation at 1110 nm is:
\begin{align}
\label{eq:etta_1110}
   |\eta_{lk}|&= 10^{4} \times \\ \nonumber
   &
\begin{bmatrix} 
    0.2  &  0.6  &  0.6  &  0.5  &  0.1  &  0.6  &  0.5  &  4.4  &  0.1\\
    0.6  &  0.5  &  0.3  &  2.7  &  0.5  &  0.8  &  7.0  &  0.5   &  37.9\\
    0.6  &  0.3  &  2.7  &  0.1  &  0.9  &  8.8  &  1.1  &  18.7  &  0.6\\
    0.5  &  2.7  &  0.1  &  14.2  &  8.2  &  1.1  &  5.9  &  1.2  &  18.5\\
    0.1  &  0.5  &  0.9  &  8.2  &  0.5  &  0.2  &  0.8  &  0.1  &  118\\
    0.6  &  0.8  &  8.8  &  1.1  &  0.2  &  0.6  &  0.5  &  146  &  0.5\\
    0.5  &  7.0  &  1.1  &  5.9  &  0.8  &  0.5  &  149  &  0.3  &  33.1\\
    4.4  &  0.5  &  18.7  &  1.2  &  0.1  &  146  &  0.3  &  36.4  &  0.1\\
    0.1  &  37.9  &  0.6  &  18.5  &  118  &  0.5  &  33.1  &  0.1  &  22.8\\
\end{bmatrix} 
\end{align}

The nonlinear coupling term for the pump excitation at 1125 nm is:
\begin{align}
\label{eq:etta_1125}
   |\eta_{lk}|&= 10^{4} \times \\ \nonumber
   &
\begin{bmatrix} 
    0.4  &  0.6  &  6.5  &  0.6  &  0.1  &  1.1  &  0.7 &   8.1 &   0.2\\
    0.6  &  5.8  &  0.3  &  7.0  &  2.5  & 1.0  &  6.3  &  0.7  &  6.7\\
    6.5  &  0.3  &  8.3  &  0.1  &  1.1  &  9.3  &  1.1  &  8.7  &  0.8\\
    0.6  &  7.0  &  0.1  &  2.7  &  15.5  &  1.2  &  12.5  &  1.4 &   9.2\\
    0.1  &  2.5  &  1.1  &  15.5 &   0.3  &  0.3  &  7.2  & 0.1 &   2.6\\
    1.1  &  1.0  &  9.3  &  1.2  &  0.3  &  7.3  &  0.6  &  1.9  &  0.5\\
    0.7  &  6.3  &  1.1  &  12.5  &  7.2 &   0.6 &   1.8  &  0.5 &   1.5\\
    8.1  &  0.7  &  8.7  &  1.4  &  0.1  &  1.9  &  0.5  &  2.0  &  0.3\\
    0.2  &  6.7  &  0.8  &  9.2  &  2.6  &  0.5  &  1.5  &  0.3  &  3.0\\
\end{bmatrix} 
\end{align}
The eigenvalues at these two wavelengths are shown in Fig.~\ref{fig:NanoOPO_Eigenvalues_1110_1125}. It is seen that at the threshold, since the signal and idler frequency separation is large, the parametric gain is low. However, when a phase transition from non-degenerate to degenerate  case occurs, the gain boosts rapidly.

\section{The evolution of supermodes}

The supermodes are the eigenvectors of $\pmb{\mathcal{H}} (b)$. The eigenvectors for all eigenvalues are displayed in Fig.~\ref{fig:NanoOPO_Eigenvectors_all}. The odd and even numbers correspond to the signal and idler modes, respectively. The eigenvectors, $a_{k,m}^{(a)}$, corresponding to the eigenvalues illustrated in Fig. 4 in the main text is displayed in Fig.~\ref{fig:NanoOPO_Exceptional_Points}. 

\section{Quasi-normal mode formulation}

The expansion of fields in a 3D resonator to multi-polar Mie resonances, which we have used in the main text, satisfies orthogonality and completeness only for spherical and non-dispersive structures. Hence, it cannot be applied to the general case of a resonator with an arbitrary shape. For a dispersive material, the conventional form of source-free Maxwell's equations cannot be written as a standard linear eigenproblem \cite{yan2018rigorous}. Recently, Lorentz reciprocity theorem \cite{lalanne2018light,sauvan2013theory} has been proposed to find the linear response of arbitrarily shaped plasmonic and dielectric resonators composed of a material with single-pole Lorentz dispersion in the form of $\varepsilon(\omega)=\varepsilon_{\infty}\left(1-\frac{\omega_p^2}{\omega^2-\omega_0^2+i\gamma\omega}\right)$. In this approach, two auxiliary fields are introduced: the polarization, $\vec{P}=-\varepsilon_{\infty}\left(1-\frac{\omega_p^2}{\omega^2-\omega_0^2+i\gamma\omega}\right)\vec{E}$, and the current density, $\vec{J}=-i\omega\vec{P}$, to reformulate the Maxwell's equation in a linear form \cite{lalanne2018light}: 
\begin{align}
\begin{bmatrix} 
    0  &  -i\mu_0^{-1}\nabla\times  &  0  &  0\\
    i\varepsilon_\infty^{-1}\nabla\times  &  0  &  0  &  -i\varepsilon_\infty^{-1}\\
    0  &  0  &  0  &  i\\
    0  &  i\omega_p^2\varepsilon_\infty  &  -i\omega_0^2  &  -i\gamma\\
\end{bmatrix} 
\begin{bmatrix} 
     \vec{E}_m\\
     \vec{H}_m\\
     \vec{P}_m\\
     \vec{J}_m\\
\end{bmatrix} 
=\omega_m
\begin{bmatrix} 
     \vec{E}_m\\
     \vec{H}_m\\
     \vec{P}_m\\
     \vec{J}_m\\
\end{bmatrix}
.
\end{align}
By applying proper boundary conditions \cite{yan2018rigorous}, this approach can be used to precisely find quasi-normal modes for an arbitrarily shaped 3D resonator. 
Beside the quasi-normal modes, this approach can find a continuum of background modes which depends on the boundary conditions, and can form a complete basis combined with quasi-normal modes.

Because of the low Q nature of the background mode, their contribution on the OPO threshold is negligible. However, they can change the field distribution of supermodes and their spectral response above the threshold. The connection between the quasi-normal modes and the density of states, $\rho(\omega)$, has been discussed in previous works \cite{sauvan2013theory, muljarov2016exact}. 

If we have a continuum of states, the summation in Eq.~\ref{eq:ND_signal_multimode} is converted to an integral form as:
\begin{align}
\label{eq:ND_signal_multimode_continuum}
 \frac{d}{dt}a_l^{(s)} =\left(i\delta\omega_l^{(a)}-\frac{\alpha_l^{(a)}}{2}\right)a_l^{(s)}+ib\int d\omega\rho(\omega) \eta_{l\omega}a^{(i)*}_\omega.
\end{align}
Since the effect of low-Q background modes are negligible, to simplify the numerical calculations, we can discretize Eq.~\ref{eq:ND_signal_multimode_continuum} around the quasi-normal modes:
\begin{align}
\label{eq:ND_signal_multimode_continuum_discritized}
 \frac{d}{dt}a_l^{(s)} =\left(i\delta\omega_l^{(a)}-\frac{\alpha_l^{(a)}}{2}\right)a_l^{(s)}+ib\sum_k\int d\omega\rho_k(\omega) \eta_{lk}a^{(i)*}_k,
\end{align}
where $\rho_k(\omega)$ is the density of states around the resonant frequency of the $k^{th}$ quasi-normal mode of the resonator.

\bibliography{references}

\end{document}